\documentclass[10pt]{bmc_article}    

\usepackage{cite} 
\usepackage{url}  
\usepackage{ifthen}  
\usepackage{multicol}   
\usepackage{epsfig} 
\newboolean{publ}

\newenvironment{bmcformat}{\begin{raggedright}
    \baselineskip20pt\sloppy\setboolean{publ}{false}}{\end{raggedright}
  \baselineskip20pt\sloppy}

\begin{document}
\begin{bmcformat}


  \title{Identification of candidate regulatory sequences in mammalian 3' UTRs
    by statistical analysis of oligonucleotide distributions} 
  

  \author{Davide Cor\`a$^{1}$%
    \email{Davide Cor\`a - cora@to.infn.it}
    \and
    Ferdinando Di Cunto$^{2}$%
    \email{Ferdinando Di Cunto - ferdinando.dicunto@unito.it}
    \and 
    Michele Caselle$^{1}$%
    \email{Michele Caselle - caselle@to.infn.it}
    and 
    Paolo Provero\correspondingauthor $^{2}$%
    \email{Paolo Provero\correspondingauthor - paolo.provero@unito.it}%
  }


  \address{%
    \iid(1)Dept. of Theoretical Physics, University of Turin and INFN,
    Turin, Italy\\ 
    \iid(2)Molecular Biotechnology Center and 
    Dept. of Genetics, Biology and Biochemistry, University of Turin, Italy }%

  \maketitle


  \begin{abstract}
    \paragraph*{Background:} 3' untranslated regions (3' UTRs) contain binding
    sites for many regulatory elements, and in particular for microRNAs
    (miRNAs). The importance of miRNA-mediated post-transcriptional regulation
    has become increasingly clear in the last few years.

    \paragraph*{Results:} We propose two complementary approaches to the
    statistical analysis of oligonucleotide frequencies in mammalian 3' UTRs
    aimed at the identification of candidate binding sites for regulatory
    elements. The first method is based on the identification of sets of genes
    characterized by evolutionarily conserved overrepresentation of an
    oligonucleotide. The second method is based on the identification of
    oligonucleotides showing statistically significant strand asymmetry in
    their distribution in 3' UTRs.

    \paragraph*{Conclusions:} Both methods are able to identify many
    previously known binding sites located in 3'UTRs, and in particular seed
    regions of known miRNAs. Many new candidates are proposed for experimental
    verification.

  \end{abstract}

  \ifthenelse{\boolean{publ}}{\begin{multicols}{2}}{}



    \section*{Background}

    The pathway leading from a gene sequence to the corresponding protein is
    organized in several steps, all subject to specific regulatory events:
    from the control of transcription initiation to complex post-translational
    events that ultimately regulate the fate of the protein product.
    Increasing evidence indicates that 3' UTRs (3'-untranslated regions) of
    mRNAs contain different types of short sequence elements playing an
    important role in the post-transcriptional control of gene expression,
    regulating mRNA stability, localization and translation efficiency
    \cite{demoor:2005}.

    In particular, a class of small RNAs called micro-RNAs mediate a
    widespread mechanism of post-transcriptional regulation. Its importance
    has been clarified in the last few years (reviewed in \cite{he:2004} and
    \cite{du:2005}).  MicroRNAs (miRNAs) are $\sim$ 22nt small non-coding RNAs
    which negatively regulate gene expression at the post-transcriptional
    level, in a wide range of organisms. They are involved in many different
    biological functions, including, in animals, developmental timing,
    pattern formation and embryogenesis, differentiation and organogenesis,
    growth control and cell death. MicroRNAs are also known to be relevant to
    human diseases \cite{alvarez-garcia:2005, calin:2006}.

    Mature and active miRNAs are thought to be produced from longer $\sim$
    200nt RNA precursors characterized by imperfect stem-loop structures.
    These long RNA precursors (pri-miRNAs) are transcribed by RNA polymerase
    II from particular loci on the genomic DNA, usually called microRNA genes
    \cite{lee:1993, wightman:1993, lee:2001, lee:2004}.  In animals,
    pri-miRNAs undergo a series of transformations to become mature miRNAs.
    The latter need to be coupled with a special protein complex called
    RNA-Induced Silencing Complex (RISC) to become effective as gene
    regulators\cite{lee:2003, bernstein:2001, schwartz:2003, khvorova:2003}.

    Even though the precise mechanism of action of the miRNA/RISC complex is
    not very well understood, the current paradigm is that miRNAs are able to
    negatively affect the expression of a target gene via mRNA cleavage or
    translational repression \cite{ambros:2004, lim:2005}, after antisense
    complementary base-pair matching to specific target sequences in the 3'
    UTR of the regulated genes. In plants, miRNAs usually have perfect or near
    perfect complementarity to their mRNA target, whereas in animals the
    complementarity is restricted to the 5' regions of the miRNA, in
    particular requiring a "seed" of 7 nucleotides, usually (but not always)
    from nucleotides 2 to 8 \cite{lai:2002, lewis:2003, john:2004, lewis:2005,
      xie:2005, chan:2005, grun:2005}.

    To date, hundreds of miRNAs have been annotated in the genomes of various
    metazoan organisms together with some of their targets. Each miRNA can
    regulate between a few and a few hundred genes. In particular, more than
    400 miRNA genes have been identified in the human genome and up to one
    third of the human protein-coding genes is currently believed to be
    regulated by them \cite{lewis:2003, john:2004,lewis:2005, krek:2005,
      berezikov:2005, xie:2005, bentwich:2005, robins:2005, chan:2005,
      griffiths:2006}.  The miRNA binding site is often overrepresented in the
    3' UTR sequence of the target gene.  Regulation by miRNA is likely a
    combinatorial mechanism, meaning that a certain mRNA can be under the
    control of many different miRNAs \cite{krek:2005}.

    miRNAs show interesting evolutionary properties between different species.
    Indeed, up to one third of the miRNAs discovered in \emph{C. elegans} have
    a human ortholog. On the other hand, species-specific miRNAs exist and, in
    particular, it has been established that primates have their own class of
    miRNA genes \cite{berezikov:2005}.

    Several computational approaches have been developed in the last four
    years to investigate this regulatory mechanism (see \cite{rajewsky:2006}
    for a recent review). In particular, computational approaches were
    suggested for following problems:
    \begin{itemize}
    \item identification of miRNA genes.
    \item identification of genes regulated by miRNAs.
    \item description of the regulatory network established by this class of
      molecules.
    \end{itemize}

    Most computational methods proposed to identify miRNA targets are based on
    some of the following elements:
    \begin{itemize}
    \item evolutionary conservation of miRNAs and their binding sites between
      species.
    \item use of the Watson - Crick perfect or imperfect pairing between 3'
      UTRs and the miRNAs seeds.
    \item enrichment of miRNA binding sites in 3' UTRs.
    \item use of RNA secondary structure information.
    \end{itemize} 

    Important aspects of the effect of miRNAs on the mammalian transcriptome
    were established in Ref. \cite{farh:2005}. In particular the following
    points will be important for our analysis:
    \begin{itemize}
    \item thousands of mammalian genes are under selective pressure to
      maintain miRNA binding sites in their 3' UTRs
    \item evolutionary conservation of the binding sites is a powerful tool to
      identify biologically relevant sites, not because non-conserved sites
      are unable to mediate repression, but because they tend to appear in
      genes which are not co-expressed with the corresponding miRNA
    \item mRNAs with a miRNA binding site are systematically depleted in the
      tissues where the miRNA is expressed compared to mRNAs with the same
      expectation for having sites, taking UTR length and nucleotide
      composition into account
    \end{itemize}

    In this work we present two new methods for the identification of miRNA
    binding sites, or, more generally, of regulatory sequences located in the
    3' UTR of the mRNA. The methods are based on the frequency distribution of
    oligonucleotides in the 3' UTRs and on evolutionary conservation, and
    specifically on two working hypotheses:
    \begin{enumerate}
    \item {\it conserved overrepresentation}: binding sites occur on regulated
      genes more often than expected by chance, and such overrepresentation is
      conserved in orthologous genes of closely related species
    \item {\it strand asymmetry}: binding sites appear in 3' UTRs, taken as a
      whole, more often than their reverse complement.  The rationale is that
      if many genes are under positive selective pressure to maintain binding
      sites in their 3' UTR, this should cause a global overrepresentation of
      the binding sites compared to their reverse complements which are not
      subjected to such positive pressure.
    \end{enumerate}

    The main novelty of the conserved overrepresentation method is that both
    the length of 3' UTR sequences and their global nucleotide composition is
    taken into account when determining whether a certain oligonucleotide is
    overrepresented in a given 3' UTR. The relative merit of this approach
    compared to the ones based simply on conserved instances of a motif in 3'
    UTRs (such as {\it e.g } Refs. \cite{xie:2005} and \cite{chan:2005}) will
    be assessed {\it a posteriori} by comparing our results to those obtained
    with such methods.

    On the contrary, strand asymmetry has not been used before as a method to
    identify regulatory elements in Eukaryotes (in both Refs. \cite{xie:2005}
    and \cite{chan:2005} strand asymmetry is referred to differences in
    conservation scores between a motif and its reverse complement, while our
    definition refers to a single genome). Strand asymmetry was used in
    \cite{bell:1998, lao:2000} to identify elements regulating recombination
    in bacteria.
    
    More precisely, the assumption we make is the existence of a {\it
      statistical correlation} between conserved overrepresentation and strand
    asymmetry of an oligonucleotide on one hand and its role as a regulatory
    element on the other. This does not imply that either property is
    necessary for an oligonucleotide to be a regulatory element.  For example,
    it is known that in several cases a single instance of the binding site
    (which in most cases is not sufficient to determine statistical
    overrepresentation) is enough to allow the miRNA to exert its regulatory
    action. Moreover, since our methods work on fixed words, they rely on
    perfect matching between miRNA and 3' UTR of the target gene, while it has
    been shown \cite{didiano:2006} that ``wobbly'' G-U pairing does not
    compromise the regulatory action of the {\it C. elegans} miRNA {\it lsy-6}
    on one of its targets.  On the other hand, {\it in vivo} experiments
    \cite{brennecke:2005} on the effectiveness of regulation by a {\it
      Drosophila} miRNA showed that (1) mismatches in the seed (positions 2 to
    8) significantly reduced the effectiveness of the regulation while (2) the
    presence of multiple copies of the seed in the 3' UTR strongly increased
    it.

    Given these results, and since few miRNA/mRNA interactions have been
    experimentally validated and studied, while many thousands are believed to
    exist, we conclude that the only way to test whether these statistical
    correlations exist is {\it a posteriori}. Therefore, we will select the
    oligos displaying conserved overrepresentation and/or strand asymmetry and
    show that they overlap in a statistically significant way with the binding
    sites of known regulatory elements acting on the 3' UTR, and in
    particular with seed regions of known miRNAs.

    The two methods must be considered as complementary, as they are expected
    to identify different categories of binding sites. A binding site will
    show strand asymmetry if it is shared by a large number of 3' UTRs.
    Therefore, we expect the strand asymmetry method to be effective in
    identifying sites involved in the regulation of 3'-processing, and the
    binding sites of miRNAs with a large number of target genes. On the other
    hand, the conserved overrepresentation method is expected to identify those
    binding sites that tend to appear repeated in the 3' UTR of their targets,
    and obviously only the ones that are conserved between human and mouse.
    Since we do not use any information about known miRNAs, the methods we
    develop are potentially able to identify the binding sites and target
    genes of both known and unknown miRNAs. A flow-chart of the methodology is
    depicted in Figure 1.

    
    \section*{Results and Discussion}

    We analyzed repeat-masked 3' UTR sequences of human and mouse genes using
    two different methods, both based on the statistical properties of
    oligonucleotide frequencies:
    \begin{itemize}
    \item {\it Conserved overrepresentation.} We constructed, separately for
      human and mouse, sets of genes sharing overrepresented oligonucleotides
      (``oligos'' in the following). We then selected the oligos for which the
      human and mouse sets contained a statistically significant fraction of
      orthologous genes.
    \item {\it Strand asymmetry.} We identified the oligos showing
      statistically significant strand asymmetry in their frequency
      distribution, that is a difference in frequency between the oligo and
      its reverse complement.
    \end{itemize}

    The first method is based on evolutionary conservation without resorting
    to any alignment procedure: this approach was termed ``network-level
    conservation'' in \cite{elemento:2005} and applied to 3' UTR regulatory
    elements of flies and worms in \cite{chan:2005}. The main difference
    between our "conserved overrepresentation" method and the one used in
    \cite{chan:2005} is our use of statistical overrepresentation rather then
    mere presence of an oligonucleotide in the 3' UTR region. On the other
    hand the ``strand asymmetry'' approach can be applied to a single genome,
    and the evolutionary conservation of the results can be checked {\it a
      posteriori}. We now turn to a more detailed description of the two
    methods and of the results found, leaving the more technical details for
    the ``Methods'' section.

    \subsection*{Conserved overrepresentation}
    \subsubsection*{Sets of genes sharing an overrepresented oligo in their 3'
      UTR region}

    For each oligonucleotide $w$ we constructed, separately for human and
    mouse, the set of genes such that $w$ is overrepresented in the 3' UTR.
    The analysis was performed for oligo length between 5 and 8, but in the
    following we will concentrate on 7-mers. The definition of
    overrepresentation we adopted is that originally introduced in
    \cite{vanhelden:1998} in the context of promoter analysis. Briefly (see
    Methods for details about statistical procedures), for all $w$ we computed
    the overall frequency $f(w)$ in all 3' UTR regions, and we selected the
    genes in which the number of occurrences of $w$ is significantly higher
    than expected based on $f(w)$. Statistical significance was determined
    with a binomial distribution, and a cutoff on P-values equal to 0.01.

    This procedure is identical to the one we had previously used to identify
    candidate transcription factor binding sites in upstream sequences
    \cite{caselle:2002, cora:2004, cora:2005}, except that in the present case
    oligos which are the reverse complement of each other are not counted
    together, since we are specifically looking for regulatory elements
    located on single-stranded mRNA. Notice also that, as in the case of
    upstream regions, no correction for multiple testing was introduced in the
    construction of the sets; indeed no significance was attributed to
    overrepresentation by itself (and hence to the sets of genes): only sets
    showing significant evolutionary conservation between human and mouse, as
    described below, were selected as candidate binding sites.

    \subsubsection*{Conservation of overrepresentation}

    We then proceeded, for each oligo $w$, to examine the corresponding sets
    of human and mouse genes looking for enrichment in orthologous genes.
    Denoting by $S_H(w)$ and $S_M(w)$ such sets, we counted how many genes in
    $S_H(w)$ had an orthologous gene in $S_M(w)$, and we determined with the
    exact Fisher's test whether there was a statistically significant
    enrichment of pairs of orthologs. The Bonferroni correction for
    multiple testing was applied to these P-values, taking into account the
    number of sets compared.

    Concentrating on oligos of length 7, we obtained 465 oligos with
    evolutionarily conserved overrepresentation in 3' UTR regions [see
    Additional file 1].  These are the candidate binding sites obtained by our
    first method.

    \subsection*{Strand asymmetry}
    
    \subsubsection*{Using a Markov chain to construct null-model sequences}

    We generated a list of oligos $w$ significantly more frequent than their
    reverse complement $\bar{w}$. In principle also the oligos with
    significantly {\it lower} frequency than their reverse complements (which
    are simply the reverse complements of those included in our list) could be
    conjectured to be of biological relevance, since many genes are thought to
    be under selective pressure to avoid developing miRNA binding sites
    \cite{farh:2005}. However these oligos turn out not to be significantly
    over-represented among seed regions of known miRNAs (data not shown).

    The analysis is non trivial due to the peculiar nucleotide frequency
    distribution of 3' UTR regions, which is reported for our sequences in
    Table 1 (in agreement with previously reported results
    \cite{shabalina:2003}). The large difference between the frequencies
    of A and T implies that the simplest possible way of analyzing oligo
    strand asymmetry would be misleading: compared to a random sequence, 3'
    UTR regions would show strong overrepresentation of oligos with many T's
    compared to their reverse complement, but this would be better explained
    by the peculiar nucleotide frequencies rather than by functional
    significance.

    A more sophisticated model is therefore needed to take into account the
    nucleotide frequencies, and possibly other peculiarities of 3' UTR
    regions, for example in terms of the frequencies of short oligos. The
    natural solution is to use a Markov chain to construct a model sequence
    reproducing the oligo frequencies of the sequences under study up to a
    certain length, and then compare the distribution of longer oligos in the
    actual sequence to the model sequence. We generated 100 sets of model
    sequences, separately for human and mouse, reproducing the experimental
    oligo frequencies up to length 4. Each set of model sequences contained as
    many sequences as the real set of 3' UTR sequences with the same average
    length.

    \subsubsection*{Strand-asymmetric oligos are candidate binding sites}  

    The sequences generated by the Markov chain were used as a null
    model for the actual sequences: for each oligo $w$ we considered the
    quantity $a(w) = f(w) - f(\bar{w})$ where $f(w)$ is the frequency of $w$
    and $\bar{w}$ is its reverse complement. We computed a P-value for the
    strand asymmetry of $w$ based on the  assumption of normal distribution of
    $a(w)$ in the null model. A Bonferroni correction for multiple testing
    was applied to these P-values.

    214 oligos of length 7 showed strand asymmetry with Bonferroni-corrected
    P-value less than 0.01 in the human case, and 139 for the mouse.  Of
    these, 113 were in common (compared to $\sim$ 2 expected by chance):
    evolutionary conservation was thus recovered {\it a posteriori}, providing
    strong support for the biological relevance of the binding sites
    identified by the method. The lists of the 7-mers showing strand asymmetry
    in human and mouse are reported in the supplementary meterial [see
    Additional files 2 and 3 respectively].

    As a control, we compared these results with those obtained by the same
    analysis on the genomic sequence lying upstream of the transcription start
    site (TSS) of annotated genes. Since these are not transcribed, we do not
    expect in this case significant deviations from randomness in the
    distribution of strand asymmetry. Fig. 2 shows the distribution of the
    $z$-values calculated on upstream regions of length 3000 bp and the same
    distribution for 3' UTR regions. As expected, in the case of upstream
    regions the distribution is much narrower. Indeed only five 7-mers showed
    significant strand asymmetry: given the difficulty of determining the TSS
    by automated annotation systems, these cases can probably be explained by
    the erroneous inclusion of sequence fragments that are actually
    transcribed.

    \subsection*{Overlap between the candidate binding sites found by the two
      methods}
    
    Of the 214 7-mers identified by strand asymmetry in human, 78 were
    also identified by conserved overrepresentation (compared to $\sim
    6$ expected by chance in the null hypothesis in which these 7-mers
    are drawn at random from all 16384 possible 7-mers - P-value $5.6
    \cdot 10^{-66}$ from Fisher's exact test). Of the 113 7-mers
    displaying strand asymmetry in both organisms, 59 were also
    identified by conserved overrepresentation (expected $\sim 3.2$ -
    P-value $2.3 \cdot 10^{-61}$).

    It is worth noting explicitly that conserved overrepresentation and strand
    asymmetry as defined in this work are largely independent properties of
    $n$-mers. While the overrepresentation of binding sites on target mRNAs is
    expected to increase the positive selection which is at the origin of
    strand asymmetry, it is not overrepresentation itself that was used to
    select the candidate $k$-mers, but its conservation (virtually all
    $k$-mers are overrepresented according to our definition in at least some
    genes).

    Thus the fact that a significant number of 7-mers were identified by both
    methods is an important argument in favor of their potential biological
    relevance. However, as shown below, each method is independently able to
    identify a statistically significant fraction of seed regions of known
    human miRNAs and binding sites of other regulatory elements; therefore
    also oligos identified by only one method can be considered as candidates
    for experimental verification.
    
    \subsection*{Comparison with seed regions of known miRNAs}

    We compared the 7-mers identified by our two methods with a list of 1017
    7-mers, representing the seed regions of known human miRNAs. The results
    of such comparison ({\it i.e.} the number of positive matches) are shown
    in Table 2 for the lists of oligos determined by the two methods and
    various combinations of them [the matching 7-mers and the corresponding
    miRNAs are listed in Additional files 4 and 5]. P-values were
    obtained with the hypergeometric distribution, {\em i.e.} in the null
    hypothesis in which both the seed regions of known miRNAs and our
    candidates are represented by randomly chosen 7-mers. The use of this null
    hypothesis was justified by producing 1000 lists of 1017 7-mers each,
    randomly generated with a Markov chain reproducing the same dinucleotide
    distribution as the true seed regions, and computing the mean number of
    matches with our candidates, which turned out to be in agreement with the
    prediction of the hypergeometric distribution. The small P-values
    demonstrate the effectiveness of the method.

    In the following discussion we will concentrate on the 59 7-mers
    identified by all methods. This choice maximizes the specificity of the
    method: indeed we will show that the vast majority of these 7-mers can be
    recognized as previously known regulatory elements. To increase
    sensitivity, and thus to obtain larger lists of {\it new} candidate
    regulatory elements one can consider the 7-mers identified by each single
    method separately: the P-values presented in Table 2 demonstrate that the
    single methods taken separately have significant predictive power.
    
    \subsection*{Identification of seed regions of known miRNAs and of binding
      sites of other 3' UTR regulatory elements}

    While 14 of the 59 7-mers correspond to known miRNA seed regions (shown
    in Table 3), other known functional elements can be recognized among them
    (see Table 4).  A particularly prominent signal that we would expect to
    see is the Poly-A signal {\bf AATAAA}. Considering both perfect matches
    and ``side matches'' where 5 out of the 6 bases of the PolyA signal match
    either side of our oligo, 12 of our oligos were thus identified as PolyA
    signals. Also the variant {\bf ATTAAA} of the Poly-A signal can be
    recognized in one of the 7-mers.

    Another well known element that we  expect to find are the AU-rich
    elements (ARE), which have been recently linked to miRNAs in triggering
    mRNA instability \cite{jing:2005}, with consensus sequences {\bf
      ATTTA}. We found three of these elements among our entries.  Among the
    most interesting non-miRNA related identifications are nine PUF
    (Pumilio-FBF protein family) elements (see for instance
    \cite{wickens:2002} for a review). The consensus sequence is in this case:
    {\bf TGTANATA}. We considered both perfect matches and ``side matches''
    defined as in the case of PolyA signals but requiring a match of length 6.

    Six of our 7-mers match the CPE element {\bf TTTTAT} allowing for one
    mismatch. This element is involved in cytoplasmic polyadenylation (see
    \cite{richter:1999} for a review).  Note that {\bf TGTAN} is also the
    consensus sequence for the binding site of the CFI$_{\rm m}$ protein,
    responsible for a non-canonical mechanism of polyA site recognition
    \cite{venkataraman:2005}. Five 7-mers, including three previously
    recognized as PUF sites and one recognized as a CPE, match this consensus.

    More generally T-rich elements can  probably be recognized as the binding
    sites of CstF, a known co-factor of the poly-A site binding protein CPSF
    \cite{macdonald:1994}, or of Fip1 \cite{kaufmann:2004}.  However such
    elements appear, in our list, almost exclusively in the combinations {\bf
      TATTTT} or {\bf TTTTNT}. The first of these elements was also identified
    computationally in Ref.\cite{hu:2005}, while the second could perhaps be
    interpreted as a variant CPE element, and is marked as such in the table.

    All the known regulatory elements cited above, with consensus sequences
    shorter than 7, were correctly reproduced by at least one of the methods
    applied to 5 and 6-mers, with the exception of the alternative
    polyadenylation site {\bf ATTAAA}. The complete lists of 5-mers and 6-mers
    identified by the two methods are available in the supplementary material
    [see Additional files 8, 9, 10 and 11].

    \subsubsection*{Comparison with the binding sites of putative miRNAs}

    The miRNAMap site \cite{hsu:2006} provides lists of putative miRNAs for
    many organisms, including human, obtained using RNAz \cite{washietl:2005},
    a tool for the prediction of non-coding RNA structures. For 464 of the
    human putative miRNAs thus obtained, miRNAMap also includes a putative
    mature miRNA sequence, obtained through a machine learning method. Since
    these candidate miRNAs were obtained from a computational approach which
    is completely independent from ours, as it makes no reference to the
    sequence of the miRNA targets, the comparison of our candidate miRNA
    binding sites to the seeds of these putative miRNAs provides a further
    test of our procedure.

    We proceeded exactly as we did for known miRNAs, and obtained a list of
    939 7-mers to be compared with our candidates. Also in this case we
    obtained a satisfactory degree of overlapping, detailed in Table 5 [the
    matching 7-mers and the corresponding putative miRNAs are listed in
    Additional files 6 and 7].  The overrepresentation of matches survives
    also if we remove from these 939 7-mers the ones that are also seed
    regions of known miRNAs (data not shown).  Finally Table 6 presents the
    same comparison made with known and putative miRNA considered together.

    \subsubsection*{Comparison with putative targets of miR-124 determined in a
      microarray experiment}
    
    Recently an integrated method combining miRNA target predictions
    and expression profiling was developed in Ref. \cite{wang:2006}. In
    particular the authors published a list of 8 genes downregulated
    following overexpression of human miR-124. Since the three
    7-mers associated to the seed region of this miRNA
    were identified by our method (GTGCCTT and TGCCTTA
    by all methods - GCCTTAA by conserved overrepresentation), we
    checked whether these 8 genes were included in the sets of human
    genes characterized by the overrepresentation of these 7-mers: indeed
    five of the eight genes reported (namely  SLC16A1, VAMP3,
    LAMC1, ATP6V0E, ACAA2) do appear in one of the three sets, thus
    confirming the relevance of oligo overrepresentation in the 3' UTR
    to the binding of miRNAs.

    \subsubsection*{Putative new cis-regulatory elements}

    Only seven of the 59 7-mers identified by our method could not be
    associated to any experimentally known binding site or miRNA seed region;
    these are listed in Table 7. This shows that the use of all the methods
    combined achieves very high specificity, while clearly sacrificing
    sensitivity.  Higher sensitivity, and thus a higher number of novel
    candidates, can be obtained by considering the predictions of each method
    separately.

    \subsection*{Comparison with other computational approaches}
    
    \subsubsection*{Conserved overrepresentation {\it vs.} conserved presence}

    The first method we proposed, based on conserved overrepresentation, can
    be compared with the method proposed in Ref. \cite{chan:2005}, in which
    the simple presence of a $k$-mer is used instead of statistical
    overrepresentation. We implemented a method as similar as possible to the
    one proposed in Ref.\cite{chan:2005} on our human and mouse sequences, as
    detailed in the Methods section. To make the comparison easier we
    considered the 465 top-ranking 7-mers, that is the same number of 7-mers
    identified by our conserved overrepresentation method.

    Of these 465 7-mers, 77 were recognized as seed regions of known miRNAs
    (defined in the same way as in the previous section), compared to 74 from
    the conserved overrepresentation method. Therefore the effectiveness of
    the two methods is similar, with a slight advantage for the one proposed in
    Ref. \cite{chan:2005}. It is important, however, to notice that the 7-mers
    identified by the two methods are not the same: only 145 7-mers are in
    common between the lists determined by the two methods.  The two methods
    have therefore comparable predictive power but give markedly different
    results, and can thus be expected to complement each other.

    It is interesting to look at the distribution in length of the 3' UTR
    regions of the genes identified by the two methods. Both distributions
    [see Supplementary Figures 1 and 2] are significantly different from
    the overall length distribution of 3' UTRs, but in different ways: while
    the 3' UTRs identified by our method show underrepresentation of sequences
    of length between a few hundred and about 1000 bps, the ones identified by
    conserved presence show underrepresentation of the shortest sequences.
    This difference could explain in part the relatively small overlap between
    the regulatory elements identified by the two methods. 

    In both cases, however, short UTRs tend to be underrepresented, a bias
    that has a plausible biological explanation: it has been noticed
    \cite{eisenberg:2003} that highly expressed housekeeping genes tend to
    have short UTRs, a possible explanation for this being \cite{farh:2005}
    that such genes are under selective pressure to avoid acquiring miRNA
    target sites. Therefore it is plausible to postulate that miRNA targets
    have relatively longer UTRs. Of course there is no way to determine how
    much of the bias is due to biological reasons and how much is introduced
    by the statistical methods. 

    Finally, note that this comparison should be taken with some caution since
    in Ref. \cite{chan:2005} a correction to the conservation score is
    introduced based on the length of the 3' UTR sequences, which is not
    possible to use  in our case since it requires the 3' UTRs of orthologous
    genes to be of equal length.

    \subsubsection*{Comparison with the results of Ref. \cite{xie:2005} }

    Ref. \cite{xie:2005} describes a computational approach to the
    identification of regulatory elements in both promoters and 3' UTRs based
    on comparative genomics. The main differences between this approach and
    ours are (1) the algorithm of Ref. \cite{xie:2005} is based on
    sequence alignments; (2) it uses information from four different
    genomes (3) it is based on degenerate motifs rather than fixed words (4)
    it does not take statistical overrepresentation into account. 

    To compare the results of \cite{xie:2005} to ours we considered the 72
    motifs identified in \cite{xie:2005} as miRNA-related. These 72 motifs
    arise from the clustering of 540 different 8-mers. Comparing these 540
    8-mers to the binding sites of known miRNAs with the same procedure used
    for our 7-mers, we found that 89 of them matched the seed region of a
    known miRNA ({\it i.e.} an 8-mer appearing within distance 2 from the 5'
    end of a known miRNA). While the fraction of true positives (89/540 =
    0.165) is similar to the one we achieve with conserved overrepresentation
    (74/465 = 0.159, see Table 2) and lower than the one we obtain with strand
    asymmetry (41/214 = 0.192), the statistical significance of the results of
    Ref.  \cite{xie:2005} is much higher: the same sensitivity corresponds to
    higher statistical significance since the total number of 8-mers is four
    times the total number of 7-mers. A higher degree of statistical
    significance is to be expected since information from four genomes instead
    of two (or one in the case of strand asymmetry) was used in Ref.
    \cite{xie:2005}.

    The oligomers identified by our method are significantly different from the
    ones identified in Ref. \cite{xie:2005}: out of 465 (214) 7-mers
    identified with conserved overrepresentation (strand asymmetry), 140 (69)
    match one of the 540 8-mers of Ref. \cite{xie:2005}. Also in this case,
    therefore, the two methods can be expected to complement each other.

    \subsubsection*{Comparison with the results of Ref. \cite{hu:2005} }

    The authors of Ref. \cite{hu:2005} developed a computational method for
    the identification of a specific class of {\it cis}-regulatory elements in
    3' UTR sequences, namely those involved in the regulation of mRNA
    polyadenylation. The method radically differs from ours and from the ones
    discussed above in that it is based on the position of regulatory elements
    with respect to poly-A sites. Nevertheless, there is a large superposition
    between their results and ours: most of the hexamers cited in Table 1 of
    Ref.  \cite{hu:2005} match one of the 7-mers identified by one or both of
    our methods. Only for the cis-elements AUE.1, CDE.4 and ADE.4 none of the
    three top hexamers identified in \cite{hu:2005} matched one of our 7-mers.

    \section*{Conclusions}

    We have presented two computational approaches to the identification of
    cis-regulatory elements located on mammalian 3' UTR regions. Both methods
    are based on the distribution of oligonucleotides: the first looks for
    oligonucleotides which are overrepresented in 3' UTR regions of human
    genes and their mouse orthologs. The second method relies on the
    identification of oligos displaying statistically significant strand
    asymmetry, as it should be expected for regulatory elements binding the
    mRNA of many target genes. The identification of binding sites through
    strand asymmetry is, to the best of our knowledge, the first {\it ab
      initio} method that is based on the statistical analysis of sequences
    from a single genome.

    The effectiveness of the methods is shown by several facts: 
    \begin{itemize}
    \item A significant fraction of the candidate binding sites proposed are
      complementary to the 5' end of known human miRNAs, where the binding to
      mRNA takes place in most cases
    \item The same applies to putative miRNAs found by a completely
      independent computational method \cite{hsu:2006}
    \item The two methods, while relying on statistically independent
      properties of oligonucleotide distributions, identify many common
      candidate binding sites
    \item The candidates identified through strand asymmetry show a remarkable
      degree of evolutionary conservation even if comparative genomics is not
      used {\it a priori}
    \end{itemize}
    
    Taken together, the methods identify 610 7-mers as candidate binding
    sites. The strong statistical overrepresentation of 7-mers that can be
    recognized as seed regions of known miRNAs demonstrate their
    effectiveness.  In particular a large majority of the 59 7-mers
    characterized by conserved overrepresentation and strand asymmetry in both
    human and mouse are recognizable as experimentally known binding sites, of
    both miRNA and other {\it trans}-acting elements, such as those involved
    in the regulation of polyadenylation.

    \section*{Methods}

    \subsection*{3' UTR sequences}

    3' UTR sequences were obtained from Ensembl \cite{birney:2006}, version 36 and pre-processed
    with the following steps (identical for human and mouse):
    \begin{itemize}
    \item Repeat-masked Ensembl exons marked as 3' UTR were downloaded for
      each protein-coding annotated gene. The masking parameters were left at
      the default values provided by Ensembl.
    \item The sequences thus obtained were organized in non-overlapping
      fragments in which repeat-masked parts were removed. We thus obtained,
      for each gene, the non-masked part of the 3' UTR as a series of
      non-overlapping fragments.
    \item We used BLAST to identify duplicated fragments: we constructed a
      network of fragments in which two nodes were connected if the BLAST
      E-value was less than 1e-40, and we retained only one fragment per each
      connected component. This procedure guarantees that duplicated sequences
      appear only once in our sample.
    \end{itemize}

    We obtained 45898 human 3' UTR fragments for the human case, and 33728
    mouse ones, respectively corresponding to 16800 and 13016 distinct Ensembl
    ids. The average length of human (mouse) fragments was $\sim 364$ nt
    ($\sim 352$), while each gene was on average associated with a set of
    fragments with total length $\sim 994$ ($\sim 906$)nt.

    \subsection*{Oligo overrepresentation}

    The genes, identified by their Ensembl ids, were first divided into two
    groups based on the nucleotide composition of their 3' UTR. Indeed the
    high variability of C/G content among different genomic locations could
    induce a bias in the statistical analysis of oligo frequencies if not
    taken into account. Since the classification of genome regions into
    classes based on their nucleotide composition is still somewhat
    controversial \cite{cohen:2005, costantini:2006}, we simply divided our
    genes, separately for each organism, into two groups (CG-rich and CG-poor)
    divided by the median CG content of the 3' UTR.

    We constructed, separately for each species, the sets $S(w)$ of genes such
    that the oligo $w$ is overrepresented in the 3' UTR. This was done for all
    oligos of length between 5 and 8 through the following steps:
    \begin{itemize}
    \item We computed the overall frequency $f(w)$ as the ratio
      \begin{equation}
        f(w) = \frac{N(w)}{N}
      \end{equation}
      where $N(w)$ is the number of times $w$ occurs in the collection of all
      3' UTR sequences, and $N=\sum_w N(w)$.
    \item For each gene $g$ let $n_g(w)$ be the number of occurrences of $w$
      in the 3' UTR region of $g$ (in general made of several fragments, see
      above). We computed the overrepresentation P-value as
      \begin{equation}
        P_g(w) = \sum_{k=n_g(w)}^{n_g}\left(n_g\atop k \right) f(w)^{k}
        (1-f(w))^{n_g-k} 
      \end{equation}
      where
      \begin{equation}
        n_g = \sum_w n_g(w)
      \end{equation}
      is the total number of oligos of the same length as $w$ that can be read
      in  the 3' UTR region of $g$. Self-overlapping matches of the same oligo
      were discarded \cite{vanhelden:1998}.
    \item The genes for which $P_g(w) < 0.01$ were included in the set $S(w)$.
    \end{itemize}
    
    The procedure described above was performed separately for CG-rich and
    CG-poor genes, so that overrepresentation is defined with respect to the
    appropriate background frequencies. The sets $S(w)$ computed for CG-rich
    and CG-poor genes were then joined to obtain a single set $S(w)$ for each
    organism.

    \subsection*{Overrepresentation of miRNA binding sites in experimentally
      known miRNA-3' UTR regulatory interactions} 

    To verify whether our definition of overrepresentation is useful for the
    identification of miRNA binding sites, we downloaded from miRNAMap
    \cite{hsu:2006} a list of 27 experimentally verified instances of such
    interactions in human (two of the 29 interactions listed in
    \cite{hsu:2006} involved a miRNA which is not included in the current
    version of miRNABase). Each interaction is represented by a miRNA id and a
    human gene identified by an Ensembl id. For each miRNA involved we defined
    three possible binding sites as the reverse complement of the 7-mers
    starting at position 1, 2 and 3 of the 5' end of the mature miRNA sequence
    as given miRNABase. In 10 of the 27 instances, one of these binding sites
    was indeed overrepresented, according to our definition, in the 3'
    UTR sequence of the human gene. 

    \subsection*{Conservation of overrepresentation}  An oligo $w$ has
    conserved overrepresentation if the sets of genes $S_{human}(w)$ and
    $S_{mouse}(w)$ contain a significantly larger number of orthologous
    genes than expected by chance. Pairs of human-mouse orthologous genes
    were obtained from Ensembl, selecting only orthologs defined as Unique
    Blast Reciprocal Hit so as to obtain one-to-one orthology
    relationships.

    Let $M$ be the total number of human genes represented in our sequences
    which have a mouse ortholog. Given an oligo $w$ and the set
    $S_{human}(w)$, let $m$ be the number of human genes in $S_{human}(w)$
    which have a mouse ortholog, $N$ the number of genes in $S_{mouse}(w)$
    with a human ortholog, and $n$ the number of genes in
    $S_{human}(w)$ with a mouse ortholog in $S_{mouse}(w)$. We then
    compute the P-value 
    \begin{equation}
      P = \sum_{k=n}^{m} F(M,m,N,k)
    \end{equation}
    where
    \begin{equation}
      F(M,m,N,k) = \frac{\left(m \atop k\right) \left(M - m \atop N -       
          k\right)}{\left(M \atop N\right)}
    \end{equation}

    Multiple testing was taken into account with the Bonferroni correction,
    and conserved overrepresentation was defined to be significant when the
    Bonferroni-corrected P-value was less than 0.01.

    \subsection*{Identification strand-asymmetric oligos}

    The Markov chain used to construct the model sequences has oligos of
    length 3 as its states, plus an additional state representing the end of a
    sequence (we are referring to a simple Markov chain, in which the states
    are directly accessible by the observer, and not to a hidden Markov model
    in which they are indirectly accessible through the tokens they emit).
    The inclusion of the gap state is crucial in correctly reproducing the
    oligo frequencies derived from model sequences which, as in our case, are
    heavily fragmented. Transition probabilities between states were computed
    from the actual sequences (including the transition probability from a
    3-mer to the ``end of sequence'' state). The actual sequences used were the
    fragments obtained as described above after removing the masked repeats.
    The initial state was chosen with a probability distribution reproducing
    the distribution of the initial 3-mers in the actual sequences (that is
    the transition probabilities from the ``end of sequence'' state to each
    3-mer). Each run of the Markov chain produced therefore a set of sequences
    reproducing, asymptotically, the following features of the actual set.

    \begin{itemize}
    \item
      Frequencies of all oligos of length up to 4.
    \item
      Frequency distribution of initial and final 3-mers.
    \item
      Average sequence length.
    \end{itemize}

    The same procedure was applied to determine strand-asymmetric 7-mers in
    upstream regions to produce Fig. 2.  On the set of sequences thus produced
    we computed, for each oligo $w$, the mean $\mu(w)$ and standard deviation
    $\sigma(w)$ of the quantity
    \begin{equation}
      a(w) = f(w) - f(\bar{w})
    \end{equation}
    where $f(w)$ is the frequency of oligo  $w$ and $\bar{w}$ is the reverse
    complement of $w$. The same quantity $a(w)$ was then computed for the
    actual sequences, and a $z$ value was constructed as
    \begin{equation}
      z(w) =  \frac{a(w) - \mu(w)}{\sigma(w)}
    \end{equation}
    where $a(w)$ refers now to the actual sequence. A P-value was finally
    associated to each oligo $w$ assuming a standard normal distribution of
    the $z$-values. We retained for further analysis only oligos $w$ that are
    overrepresented with respect to their reverse complement ($z > 0$ and with
    Bonferroni-corrected P-values less than 0.01).

    \subsection*{Validation with known human miRNA}

    We downloaded from the miRBase \cite{griffiths:2006} ftp site the file:
    \verb+mature.fa+  containing the mature sequences of all currently known
    miRNAs.  454 human miRNAs are annotated in release 8.2, June 2006. To
    validate our results, we compared the reverse complement of the oligos
    selected by our algorithm to the 5' end of the mature miRNAs. 

    Specifically, we extracted three ``seeds'' for each human mature miRNA
    included in miRBase, consisting in the three 7-mers starting at the first,
    second and third nucleotide on the 5' extremity of the miRNA mature
    sequence. We thus obtained 1017 distinct 7-mers which we consider as seed
    regions of known miRNAs, to be compared with the 7-mers identified by our
    method for its validation.  This validation procedure requires a perfect
    Watson-Crick complementarity between the 5' end of the miRNA and the 3'
    UTR region of its target: while this requirement can be considered too
    conservative, allowing mismatches and wobbly binding would have increased
    the number of known miRNA binding sites to include the majority of all
    possible 7-mers, making the validation of our results meaningless. On the
    other hand our choice of considering perfect matches only is justified by
    the fact that these have a significantly lower free energy than imperfect
    or wobbly ones.

    \subsection*{Conserved overrepresentation vs. conserved presence}

    We implemented on our mammalian UTR sequences a method based on conserved
    presence rather than conserved overrepresentation, as similar as possible
    to the one proposed in Ref.  \cite{chan:2005}. For each 7-mer we
    constructed the set of genes containing one or more instances of the 7-mer
    in their 3' UTR. It should be noted that in our case, as opposed to Ref.
    \cite{chan:2005}, the length of the 3' UTR regions of orthologous human
    and mouse genes are not the same: therefore it is not possible to use the
    length-dependent correction to the conservation score introduced in Ref.
    \cite{chan:2005}.
    
    The human and mouse sets thus constructed were then tested for significant
    overrepresentation of orthologous genes with the same method used for our
    overrepresentation sets. As in Ref. \cite{chan:2005} we then used the
    conservation P-values to rank the 7-mers. The 465 highest ranking 7-mers
    were then compared to the binding sites of known miRNAs with the same
    procedure adopted for the 465 7-mers identified by conserved
    overrepresentation.

    \section*{Authors' contributions}

    The four authors jointly conceived and planned the work. Most of the data
    analysis was performed by D.C. (preparation of the sequences, conserved
    overrepresentation and comparison with available experimental data) and
    P.P. (strand asymmetry). All authors participated in writing the
    manuscript, and all read and approved the final version.

    \section*{Acknowledgments}

    This work was partially supported by the Fund for Investments of Basic
    Research (FIRB) from the Italian Ministry of the University and Scientific
    Research, No. RBNE03B8KK-006. We are grateful to the anonymous referees
    for many insightful comments and suggestions.
    

    {\ifthenelse{\boolean{publ}}{\footnotesize}{\small}
      \bibliographystyle{bmc_article}  
      \bibliography{bmc_article} }     


    \ifthenelse{\boolean{publ}}{\end{multicols}}{}



  \section*{Figures}

  \subsection*{Figure 1 - Flow-chart of the proposed methodology.}

  Flow-chart of the proposed methodology: from a list of
  human 3' UTR exons to a list of putative 3' UTR regulatory elements.

  \subsection*{Figure 2 - Distribution of strand-asymmetry $z$-values in
    3' UTR and upstream regions}
  
  The distribution of the absolute value of $z$, defined in the text as a
  measure of strand asymmetry among all possible 7-mers in 3' UTR regions
  (red) and in 3000bp upstream of the TSS (grey).


  \section*{Tables}


  \subsection*{Table 1 - Nucleotide composition of 3' UTR regions.}
  \par \mbox{}
  \par
  \mbox{
    \begin{tabular}{ccc}
      \hline\noalign{\smallskip}
      & Human  & Mouse \\ 
      \hline\noalign{\smallskip}
      A & 0.2683  & 0.2638 \\ 
      C & 0.2199  & 0.2237 \\ 
      G & 0.2210  & 0.2254  \\ 
      T & 0.2908  & 0.2871  \\
      \hline 
    \end{tabular}
  }
  \par \mbox{} \par 
  The base frequencies of 3' UTR regions in human and mouse, excluding the
  masked repeats.
  \clearpage


  \subsection*{Table 2 - Comparison between the results of our computational
    approach and seed regions of known miRNAs.}

  \par \mbox{}
  \par
  \mbox{
    \begin{tabular}{|c|c|c|c|c|c|}
      \hline\noalign{\smallskip}
      Method&7-mers    &corresponding&expected &P-value\\
      &identified&to known miRNA      &by chance&\\
      \hline\noalign{\smallskip}
      $CO$ & 465  & 74 & 28.9 & $5.0\cdot 10^{-14}$\\
      $SA$ & 214  & 41 & 13.3 & $7.9\cdot 10^{-11}$ \\ 
      $CSA$& 113  & 19 & 7.01 &  $6.3\cdot 10^{-5}$  \\ 
      $CO\cap CSA$& 59  & 14 & 3.66 & $1.1\cdot 10^{-5}$ \\ 
      $CO\cup SA$& 601  & 94 & 37.3 & $4.8\cdot 10^{-17}$\\
      \hline
    \end{tabular}
  }
  \par \mbox{} \par 
  The rows indicate the computational methods as described in the text:
  $CO$=conserved overrepresentation; $SA$= strand asymmetry; $CSA$=conserved
  strand asymmetry (oligos displaying strand asymmetry in both human and
  mouse); $CO\cap CSA$ oligos identified by both $CO$ and $CSA$; $CO\cup SA$:
  oligos identified by either conserved overrepresentation or strand asymmetry
  (human) or both. The columns are (1) Number of oligos identified
  computationally; (2) Number of these matching the seed region of known human
  miRNAs; (3) number of such matches expected by chance; (4) number of
  different known human miRNAs putatively binding the oligos (many 7-oligos
  occur in the seed region of more than one miRNA) (5) P-value from exact
  Fisher test, taking into account that there are 16384 possible 7-mers 1017
  of which are seed regions of known miRNAs.

  \clearpage


  \clearpage

  \subsection*{Table 3 - 7-mers identified by all methods matching the seed
    region of known human miRNAs}

  \begin{table}[ht]
    \centering
    \begin{tabular}{cp{10cm}}
      \hline\noalign{\smallskip}
      AGCACAA&hsa-miR-218\\
      CTTTGTA$^1$&hsa-miR-524*      hsa-miR-520d*\\
      GCACTTT&hsa-miR-520d      hsa-miR-93      hsa-miR-106a    hsa-miR-520h\\
      &hsa-miR-520a     hsa-miR-520e    hsa-miR-519b    hsa-miR-20a\\
      &hsa-miR-106b     hsa-miR-372     hsa-miR-520b    hsa-miR-17-5p\\ 
      &hsa-miR-520g     hsa-miR-519c    hsa-miR-519d    hsa-miR-20b\\
      &hsa-miR-519a     hsa-miR-520c    hsa-miR-526b*   hsa-miR-519e\\
      GGTGCTA&hsa-miR-29c       hsa-miR-29b     hsa-miR-29a\\
      GTGCAAT&hsa-miR-92b       hsa-miR-367     hsa-miR-92
      hsa-miR-363     hsa-miR-32      hsa-miR-25\\
      GTGCCTT&hsa-miR-506       hsa-miR-124a\\
      GTTTACA&hsa-miR-30a-5p    hsa-miR-30b     hsa-miR-30d
      hsa-miR-30e-5p  hsa-miR-30c\\ 
      TACTGTA&hsa-miR-101       hsa-miR-199a*   hsa-miR-144   \\
      TGCAATA&hsa-miR-92b       hsa-miR-92      hsa-miR-32\\
      TGCCTTA&hsa-miR-506       hsa-miR-124a\\
      TGTTTAC&hsa-miR-30a-5p    hsa-miR-30b     hsa-miR-30d hsa-miR-30e-5p\\  
      &hsa-miR-30c\\
      TTATATT&hsa-miR-410\\
      TTGTATA$^1$&hsa-miR-381\\
      TTTGCAC&hsa-miR-19b       hsa-miR-19a\\
      \hline
      (1) also matches CFI$_{\rm m}$ binding site
    \end{tabular}
  \end{table}
  \clearpage


  \subsection*{Table 4 - 7-mers identified by all methods matching
    other known regulatory elements}
  \begin{table}[ht]
    \centering
    \begin{tabular}{cp{4cm}}
      \hline\noalign{\smallskip}
      AATAAAC&\\
      AATAAAG&\\
      ATAAAAG&\\
      ATAAAGG&\\
      ATAAAGT&\\
      ATAAATG&\\
      ATTAAAG&PolyA\\
      CAATAAA&\\
      CCAATAA&\\
      CTAATAA&\\
      GAATAAA&\\
      GCAATAA&\\
      TCAATAA&\\
      \hline
      ATTTAAG&\\
      ATTTATA&ARE\\
      TATTTAT&\\
      \hline
      GTAAATA&\\
      GTACATA&PUF\\
      TGTAAAT&\\
      TGTACAT&\\
      TGTATAT&\\
      TTGTAAA&\\
      \hline
      ATTGTAA&\\
      TTTGTAA&PUF, CFI$_{\rm m}$\\
      TTTGTAT&\\
      \hline
      TTGTATT&CFI$_{\rm m}$\\
      \hline
      TTTTGTA& CPE, CFI$_{\rm m}$\\
      \hline
      TTTTATA&\\
      TTTTGTT&\\
      TTTATAA&CPE\\
      TTTATAT&\\
      TTTTTAT&\\
      \hline
      ATATTTT&\\
      CTATTTT&\\
      GTATTTT & CstF\\
      TATTTTG &\\
      TATTTTT& \\
      TTTTTAA& \\
      \hline
    \end{tabular}
  \end{table}
  \clearpage


  \subsection*{Table 5  - Comparison between the results of our computational
    approach and seed regions of putative miRNAs listed in miRNAMap
    \cite{hsu:2006}.}

  \par \mbox{}
  \par
  \mbox{
    \begin{tabular}{|c|c|c|c|c|c|}
      \hline\noalign{\smallskip}
      Method&7-mers    &7-mers corresponding&expected &number of&P-value\\
      &identified&to putative miRNAs      &by chance&putative miRNAs&\\
      \hline\noalign{\smallskip}
      $CO$ & 465  & 76 & 26.7 & 101 & $6.7\cdot 10^{-17}$\\
      $SA$ & 214  & 46 & 12.3 & 79  & $3.9\cdot 10^{-15}$ \\ 
      $CSA$& 113  & 23 & 6.48 & 42  &  $8.8\cdot 10^{-8}$  \\ 
      $CO\cap CSA$& 59  & 15 & 3.38 & 31 & $7.8\cdot 10^{-7}$ \\ 
      $CO\cup SA$& 601  & 98 & 34.4 & 135 & $1.8\cdot 10^{-21}$\\
      \hline
    \end{tabular}
  }
  \par \mbox{} \par 
  Same as Table 2 for putative miRNAs.  The total number of 7-mers which are
  seed regions of putative miRNA is 939.
  \clearpage


  \subsection*{Table 6 - Comparison between the results of our computational
    approach and seed regions of known and putative miRNAs}

  \par \mbox{}
  \par
  \mbox{
    \begin{tabular}{|c|c|c|c|c|c|}
      \hline\noalign{\smallskip}
      Method&7-mers    &7-mers corresponding&expected &number of known
      or&P-value\\
      &identified&to known or putative miRNAs      &by chance&putative miRNAs&\\
      \hline\noalign{\smallskip}
      $CO$ & 465  & 114 & 50.3 & 217 & $1.5\cdot 10^{-17}$\\
      $SA$ & 214  & 62 & 23.1 & 180  & $1.6\cdot 10^{-13}$ \\ 
      $CSA$& 113  & 30 & 12.2 & 96  &  $2.2\cdot 10^{-6}$  \\ 
      $CO\cap CSA$& 59  & 19 & 6.38 & 77 & $8.0\cdot 10^{-6}$ \\ 
      $CO\cup SA$& 601  & 149 & 65.0 & 285 & $3.4\cdot 10^{-23}$\\
      \hline
    \end{tabular}
  }
  \par \mbox{} \par 
  r
  Same as Table 2 for known and putative miRNAs together.  The total number of
  7-mers which are seed regions of known or putative miRNA is 1772.


  \subsection*{Table 7 - 7-mers identified by all methods and not matching
    experimentally known regulatory elements or miRNA seed regions}

  \begin{table}[ht]
    \centering
    \begin{tabular}{cp{10cm}}
      \hline\noalign{\smallskip}
      AAACTTG\\
      AATCATG\\
      GACCAAA\\
      GTTATTT\\
      TATATGT\\
      TGTGAAT\\
      TTGCCTT\\
      \hline
    \end{tabular}
  \end{table}

  \clearpage

  \section*{Additional Files}

  \subsection*{Additional file 1 --- List of 7-mers displaying conserved
    overrepresentation in 3' UTR regions}
  For each oligo we report the numbers $m$, $N$ and $n$ as defined in the
  Methods section and the significance of the overrepresentation expressed as
  $-\log_{10}$ of the $P$-value (before Bonferroni correction). We also list
  for each oligo the $n$ human genes with a mouse ortholog in the
  corresponding mouse set, listing the Ensembl id, gene symbol, Entrez gene id
  and description.

  \subsection*{Additional file 2 --- 7-mers displaying strand asymmetry in
    human 3' UTR regions}
  For each oligo we report the $z$-value as defined in the text

  \subsection*{Additional file 3 --- 7-mers displaying strand asymmetry in
    mouse 3' UTRs}
  As in additional file 2.

  \subsection*{Additional file 4 --- 7-mers displaying conserved
    overrepresentation in 3' UTRs: validation with the seed regions of known
    miRNAs}
  For each oligo we report the known human miRNAs with seed region (defined
  as described in the text) matching the oligo.

  \subsection*{Additional file 5 --- 7-mers displaying strand asymmetry in
    human 3' UTRs: validation with seed regions of known miRNAs}
  As in additional file 4.

  \subsection*{Additional file 6 --- 7-mers displaying conserved
    overrepresentation in 3' UTRs: validation with seed regions of putative
    miRNAs}
  As in additional file 4.

  \subsection*{Additional file 7 --- 7-mers displaying strand asymmetry in
    human 3' UTRs: validation with seed regions of putative miRNAs}
  As in additional file 4.

  \subsection*{Additional file 8 --- List of 6-mers displaying conserved
    overrepresentation in 3' UTRs}
  As in additional file 1.

  \subsection*{Additional file 9 --- List of 6-mers displaying strand
    asymmetry in human 3' UTRs}
  As in additional file 2.

  \subsection*{Additional file 10 --- List of 5-mers displaying conserved
    overrepresentation in 3' UTRs}
  As in additional file 1.

  \subsection*{Additional file 11 --- List of 5-mers displaying conserved
    overrepresentation in human 3' UTRs}
  As in additional file 2.

  \subsection*{Supplementary Figure 1 --- Length distribution of 3' UTRs of
    the genes with conserved overrepresented 7-mers}
  3' UTR length of genes appearing in at least one of 465 sets selected by conserved
  overrepresentation (red) compared to all genes in our dataset (green).

  \subsection*{Supplementary Figure 2 --- Length distribution of 3' UTRs of
    the genes with conserved instances of the 7-mers selected with the method
    of Ref\cite{chan:2005}}
  The genes represented in red have at least one conserved instance of one of
  the 465 highest-ranking 7-mers identified using the method introduced in
  \cite{chan:2005}. The length distribution of their 3' UTRs is compared to
  the length distribution of all the genes in the dataset (green).

  \clearpage

\begin{figure}[h]
\centering
\fbox{
\includegraphics[height=15cm]{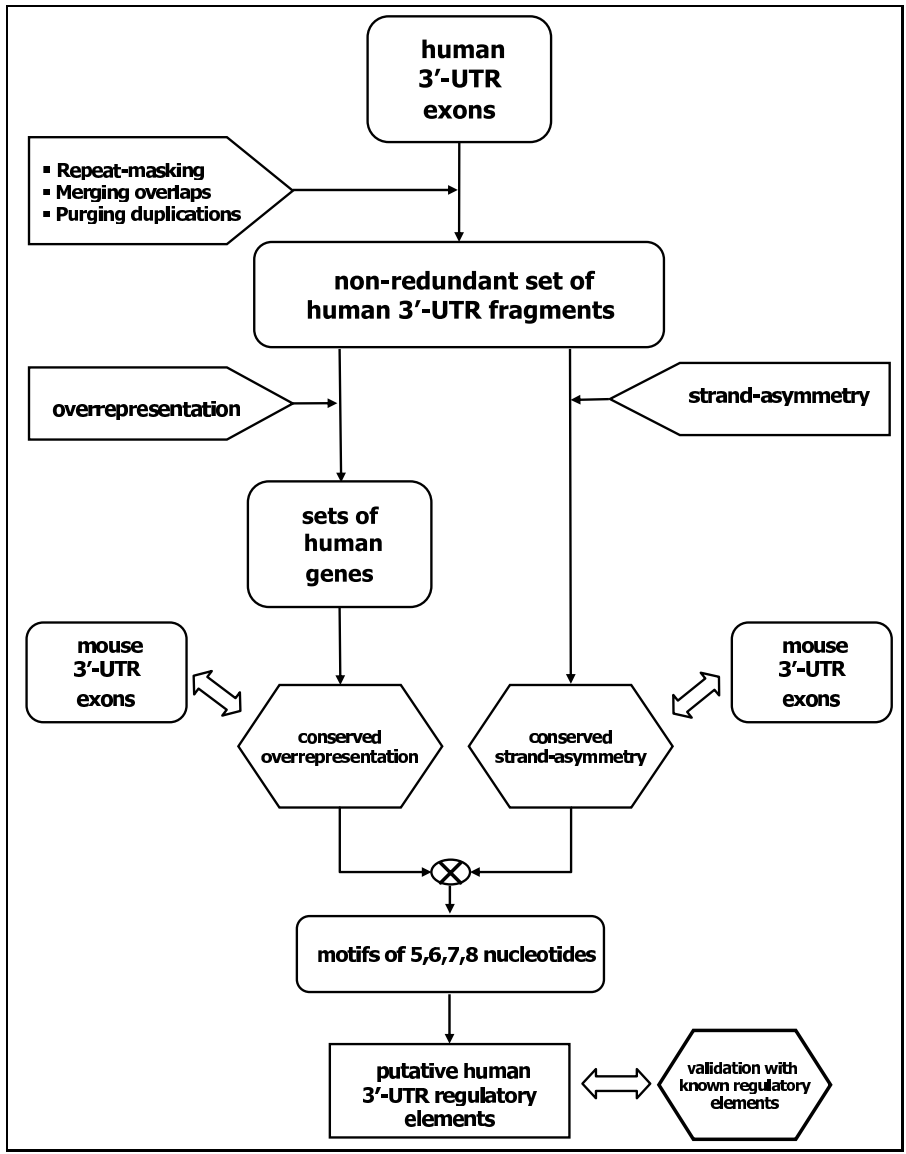}}
\caption{Flow-chart of the proposed methodology.}

\end{figure}

  \clearpage
  
\begin{figure}[h]
\centering
\fbox{
\includegraphics[height=8cm]{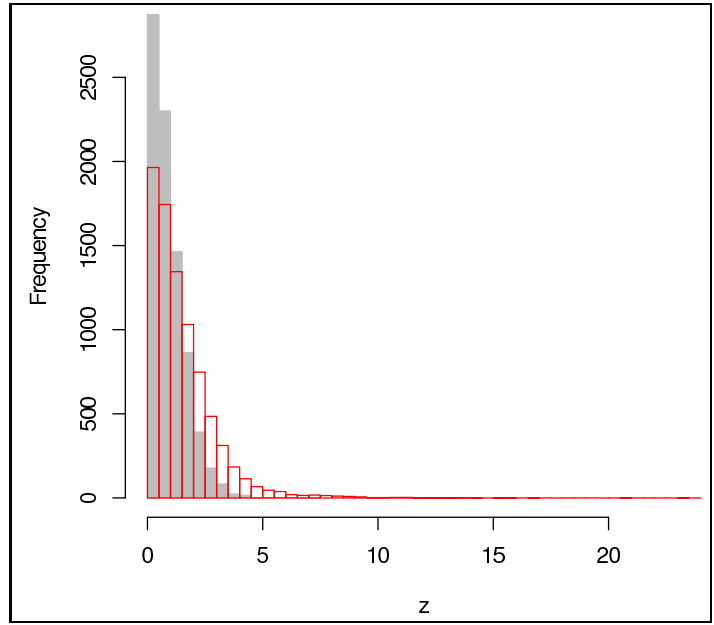}}
\caption{Distribution of strand-asymmetry z-values in 3' UTR and upstream regions.}

\end{figure}

\end{bmcformat}
\end{document}